# Coexistence of isotropic *s*-wave and extended *s*-wave order parameters in FeSe as revealed by the low-temperature specific heat

J.-Y. Lin,[1] Y. S. Hsieh,[1] D. A. Chareev,[2] A. N. Vasiliev,[3] Y. Parsons,[4] and H. D. Yang[5]

[1] *Institute of Physics, National Chiao Tung University, Hsinchu 30010, Taiwan*

[2] *Institute of Experimental Mineralogy, Chernogolovka, Moscow Region 142432, Russia*

[3] *Department of Low Temperature Physics, Moscow State University, Moscow 119991, Russia*

[4] *Department of Physics, University of California, Santa Barbara, CA 93106, USA*

[5] *Department of Physics, National Sun Yat-Sen University, Kaohsiung 804, Taiwan*

The comprehensive low-temperature specific heat $C(T)$ data identify both an isotropic *s*-wave and an extended *s*-wave order parameters coexisting in a superconducting single crystal FeSe with $T_c$=8.11 K. The isotropic gap $\Delta_0$=1.33 meV on the hole Fermi sheets and the extended s-wave gap $\Delta=\Delta_e(1+\alpha\cos2\phi)$ with $\Delta_e$=1.13 meV and $\alpha$=0.78 on the electron Fermi sheets. The extended *s*-wave is rather anisotropic but the low energy quasiparticle excitations demonstrate no sign of the accidental nodes. The coefficient $\gamma(H)$ manifesting the quasiparticle contribution to $C$ is a non-linear function of the applied magnetic field $H$ in the mixed state in accord with the anisotropic multi-order parameters.

The discovery of Fe-based high-temperature superconductivity in 2008 has been the most significant event in the field since cuprate superconductors and $MgB_2$ [1]. Alongside the multi-band electronic structure similar to that of $MgB_2$ (and in contrast to the one-band physics in cuprates) [2,3], there is keen competition between superconductivity and magnetism in Fe-based superconductors [4]. These new superconductors are likely owing to unique superconducting mechanisms, and invite rethinking on other superconductors, including cuprates. As always, the superconducting order parameter is one of the core elements toward full understanding of the superconducting mechanism. The initial proposal was that the Cooper pairs are glued by the spin excitations generated from Fermi surface nesting [5]. This pairing mechanism would lead to nearly isotropic order parameters with opposite signs on hole and electron pockets [5]. Early on, the angular resolved photoemission spectroscopy reported nearly isotropic order parameters on both hole and electron Fermi sheets consistent with the above scenario [6]. However, later nuclear magnetic resonance and thermal conductivity experiments reported existence of nodal order parameters, at least in certain 122 Fe-based superconductors [7-9]. Moreover, a recent angular resolved specific heat study of $FeSe_{0.45}Te_{0.55}$ suggests a very anisotropic order; whether it is with nodes or not remains unclear [10]. Indeed, very recent model calculations demonstrate, within reasonable physical parameter range, either *d*- or *s*-wave order parameter could be stabilized [11-13]. For the latter case, the hole bands could host an isotropic order parameter opposite to an anisotropic one on the electron bands. The critical temperature $T_c$ of FeSe scales with the density of states (DOS) at the Fermi energy $N(E_F)$ compared to those of other Fe-based superconductors [14]. Though with lower $T_c$, FeSe has the simplest structure, and this very simplicity could provide the most appropriate venue of understanding the superconducting mechanism of Fe-based superconductors [15]. In this Letter, we resolve the currently debated issue of the order parameter in FeSe by the comprehensive low-temperature specific heat data of high-quality single crystalline FeSe. For the first time, these data unambiguously prove the coexistence of an isotropic *s*-wave and a very anisotropic extended *s*-wave in an Fe-based superconductor. Furthermore, this anisotropic order parameter has no accidental nodes. The present work also better elucidates the anisotropy of both $\gamma(H)$ manifesting the quasiparticle contribution to $C$ and the upper critical field $H_{c2}$ of FeSe, both of which were rarely appropriately explored in the literature.

FeSe single crystals were grown in evacuated quartz ampoules using a $KCl/AlCl_3$ flux. The samples prepared by the $KCl/AlCl_3$ flux method were found with no impurities. The structure of tetragonal P4/nmm was demonstrated at room temperature by x-ray diffraction. The homogeneity range of tetragonal $FeSe_{1-x}$ is from FeSe$_{0.96}$ to FeSe$_{0.965}$ [16]. The present sample has the mass of 2.27 mg and its composition is FeSe$_{0.963}$ as proved by an average of the measurements of the energy-dispersive X-ray spectroscopy performed on a CAMECA SX100 (15 keV) analytical scanning electron microscope. The low temperature specific heat $C(T,H)$ was measured with a $^3$He heat-pulsed thermal relaxation calorimeter in the temperature range from 0.5 K to 15 K by applying magnetic fields with $H\perp c$ and $H//c$ up to $H=9$ T, respectively.

Fig. 1 shows both the zero-field and the mixed state specific heat $C(T,H)$ of FeSe plotted as $C/T$ vs. $T^2$ with the magnetic field $H$ varying from 0 to 9 T. The original $C$ vs. $T$ data at $H=0$ are shown in the inset of Fig. 1. The specific heat jump associated with the superconducting transition is obvious. The superconducting transition temperature $T_c$ is 8.11 K as determined by the local entropy balance. $C/T$ approaches zero at $T=0$, as will be seen more clearly in Fig. 3. The coefficient $\gamma$ at the linear-$T$ term of $C(T)$ should be absent in the superconducting state. This absence of the so called "residual $\gamma$" indicates high quality of the present sample with a complete superconducting volume, and help to avoid many complications encountered in the previous specific heat studies on Fe-based superconductors [17]. The data in $H$ allow the normal state $C_n(T)$ to be determined more reliably. $C_n(T)$ below 10 K can be well described by the simple expression $C_n(T)=\gamma_n T+\beta T^3$ where $\gamma_n T$ is the normal electronic contribution and $\beta T^3$ represents the phonon contribution. The resultant $\gamma_n$=5.73±0.19 mJ/mol K$^2$ and $\beta$=0.421±0.002 mJ/mol K$^4$ (the corresponding Debye temperature $\Theta$ is 210 K). Both are consistent with previous results from the polycrystalline samples [18]. The superconducting electronic contribution $C_{es}(T)$ can be obtained by $C_{es}(T)=C(T)-\beta T^3$. $C_{es}(T)/T$ vs. $T/T_c$ for $H$=0 is shown in Fig. 2. The entropy conservation required for a second order phase transition is fulfilled as shown in the inset of Fig. 3. This check warrants the thermodynamic consistency for both the measured data and the determination of $C_n(T)$. By the balance of entropy around the transition, the dimensionless specific-heat jump $\delta C/\gamma_n T_c$=1.65 at $T_c$ is determined compared with the BCS value of 1.43. Without any model fitting, this value has already implied a moderate or stronger coupling in FeSe.

The data are obviously not reconcilable with the conventional $s$-wave order parameter, due to the significant quasiparticle contribution at low $T$. (For a typical example of $C_{es}(T)/T$ of a conventional superconductor, see Ref. [19].) Actually, the quasiparticle contribution to $C_{es}(T)$ at low $T$ is even larger than that of the well known two-gap superconductor MgB$_2$ [20, 21]. To elucidate the order parameter of FeSe, data in Fig. 2 were fit into various models. The cases of $d$-wave, extended $s$-wave, two-gap (S$_\pm$), and $s$+(extended $s$-wave) (s+ES) are shown in Fig. 2(a), (b), (c), (d), respectively. The order parameters used to fit the data are $\Delta=\Delta_0\cos2\phi$ for $d$-wave and $\Delta=\Delta_e(1+\alpha\cos2\phi)$ (where $\alpha$ denotes the gap anisotropy) for the extended $s$-wave. In the two-gap model, two (assumed) isotropic order parameters $\Delta_L$ and $\Delta_S$ are introduced as in the previous works [22-24]. In the $s$+ES case, $\Delta=\Delta_0$ for an isotropic $s$-wave is assumed. These four cases are all allowed from the symmetry aspect and model calculations [11-13]. Considering the whole temperature range, the $s$+ES and the two-gap models lead to more satisfactory fits of the data than either $d$-wave or extended $s$-wave does.

In order to distinguish between the $s$+ES and two-gap models, further insight into the quasiparticle excitations below 0.25 $T/T_c$ is provided by Fig. 3. Thanks to the precision of the measurements, the fitting is surprisingly selective. *Conclusively, the s+Es model leads to the most preferable fit among alternatives.* Since $\alpha=0.78$, *the results illustrate a very anisotropic order parameter existing in FeSe.* This conclusion is consistent with the very recent angular resolved specific heat measurements on $FeSe_{0.45}Te_{0.55}$ [10]. Furthermore, *the present work excludes the scenarios of the d-wave line nodes* as shown in Fig. 3, an issue which could not be concluded in the previous literature [10]. From the theoretical side, the robustness of the $s$+ES model has been confirmed very recently [25]. An $\alpha$ approaching 1 (or >1) would generate too much low energy excitations to reconcile with the present experimental data (fitting not shown).

Intriguingly, there have been two scanning tunneling microscopy (STM) studies of Fe(Se,Te) and FeSe, respectively (see Refs.[26] and [27]). These two separate works reported nodes in FeSe and no node in Fe(Se,Te), respectecely. Since the minimum of the present anisotropic gap is ~0.2 meV, the observation of no accidental node might not contradict the experimental data of STM in Fe(Se,Te). The gap maxima $\Delta_e(1+\alpha)$=2.01 meV is also close to the maxima gap ~2.2 meV in FeSe by STM. To have a quantitative comparison to the nodal scenario in Ref. [27], the zero-field $C_{es}(T)$ was fit to an isotropic gap $\Delta_0$ (representing the order parameter on the hole pockets) combined with an extended $s$-wave order parameter $\Delta=\Delta_e(\cos k_x+\cos k_y)$ which results in nodes on the electron pockets. The fitting curve was denoted as the green line in Fig. 3 which is apparently inferior to that of $s$+ES, albeit slightly better than that of the nodal $d$-wave. Consideration of $\Delta=\Delta_1\cos k_x\cos k_y+\Delta_2(\cos k_x+\cos k_y)$ for the whole unfolded Brillouin zone cannot lead to any satisfactory fit. Therefore, the scenario of accidental nodes is as unlikely in FeSe. However, if the nodes cross the portion of the Fermi surface with the low DOS, the quasiparticle contribution to $C_{es}(T)$ will be small and the nodal scenario could reconcile with the low-temperature data in Fig. 3. The resultant fitting parameters for various models are summarized in Table I. With a simplified two band model in Fig. 2(d), rather than the more realistic four band model, how much the ratio of $\gamma_s/\gamma_{ES}$=33%:67% reflects the exact physical parameters remains to be seen. However, this apparently imperfect match might undermine the effect of Fermi surface nesting on the magnetic ordering and help pairing states win over the spin density wave in FeSe.

It is noted that the recent angular resolved thermal conductivity study proposes a nodal structure in the 122 system, together with other work as mentioned above. Evidence of nodes in these pnictides might not contradict the present results, since the model calculations have demonstrated that the gap symmetry is very sensitive to the model parameters and may vary among the pnictide family [11-13]. As for the previously focused two gap model in Fig. 2(c), the values of $2\Delta_L/kT_c$ and $2\Delta_s/kT_c$ are different from those of FeTe$_{0.55}$Se$_{0.45}$ [28] but are nearly identical to those of FeSe probed by Andreev spectroscopy at 4 K [29]. The qualitative results of multigap nodeless superconductivity in FeSe were reported earlier [30], though the extended $s$-wave feature and the magnitude of the order parameters were elusive then.

Fig. 4(a) shows the mixed state linear coefficient $\gamma(H)$ as obtained from the empirical fit of the in-field data between 0.5 K and 2 K by $C(T,H)/T=\gamma(H)+aT+bT^2$ where $\gamma(H)$, $a$, and $b$ are fitting parameters. To illustrate the anisotropy in FeSe, $\gamma(H)$ with $H//c$ and $H\perp c$ ($C(T,H\perp c)/T$ data not shown) was depicted in Fig. 4(a). Previously, $\gamma(H)$ was thought to be linear with respect to $H$ for the isotropic $s$-wave pairing and proportional to $H^{1/2}$ for line nodes (see discussions in Refs. [17] and [18]). In general, $\gamma(H)$ is very similar to that of $BaFe_2(As_{0.7}P_{0.3})_2$ for $H$ up to 35 T [31]. In low fields, $\gamma(H)$ show a pronounced curvature. In high fields, $\gamma(H)$ is quasi-linear. The dashed lines are the linear fits for $H//c$ and $H\perp c$ data above $H=1$ T. Knowing that $\gamma_n=5.73$ mJ/mol K$^2$, the upper critical field can be estimated as $H_{c2,H//c}\approx13.1$ T and $H_{c2,H\perp c}\approx27.9$ T. The anisotropy in $H_{c2}$ is about 2.1. Although this ratio is moderate, FeSe is already one of the most anisotropic Fe-based superconductors [32]. For the whole field range, $\gamma(H)$ with either $H//c$ or $H\perp c$ is qualitatively in accord with $S_\pm$ or $s+ES$ [33,34]. It is likely that $\gamma(H)$ in Fig.1(d) of Ref. [33] would resemble the present results even more if the $s+ES$ was used in the calculations. Fig. 4(b) show $H_{c2}(T)$ determined by the local entropy balance with $H//c$ and $H\perp c$. For both sets of $H_{c2}(T)$, in contrast with the linear $T$ dependence near $T_c$ for conventional superconductors, the results show a pronounced positive curvature with respect to $T$. This feature generally manifests multi-gap order parameters [22]. (Other interpretations can be seen in Ref. [35].) It is noted that the anisotropic $H_{c2}(T)$ was rarely explored in the literature. The present $H_{c2}(T)$ in Fig. 4(b) shows a much more pronounced negative curvature with respect to $T$ in high fields as compared with that in Ref. [36].

Finally, $\gamma_n$ of the present work is compared with the bare $\gamma_0=2.24$ mJ/mol K$^2$ from the band structure calculations [14]. Since $\gamma_n=(1+\lambda)\gamma_0$, where $\lambda$ is the total coupling strength between quasiparticle and bosons, $\lambda=1.55$ is estimated. This coupling strength is slightly stronger than the moderate coupling. The electron-phonon coupling strength $\lambda_{ep}=0.17$ was calculated in Ref. [14]. Therefore, the quasiparticles in FeSe mainly couple with other gluons rather than phonons such as those from spin fluctuations. The value of $\lambda=1.55$ is in semi-quantitative agreement with both the magnitude of $\delta C/\gamma_n T_c$ and the energy gap from fitting in Table I. The fundamental properties of FeSe are summarized in Table II.

To conclude, the present work selects nodeless $s+$ES from several alternative theoretical models. Furthermore, other precious normal state and mixed state physical properties are revealed. The low-temperature specific heat on high quality single crystals, if well executed, is indeed a powerful tool for probing the order parameter in Fe-based superconductors.

This work was supported by National Science Council of Taiwan, ROC, under Nos. NSC98-2112-009-005-MY3 and NSC100-2112-M110-004-MY3; by grant of the President of the Russian Federation MK-1557.2011.5, by the State Contract 11.519.11.6012 of Russian Ministry of Science and Education and by the Russian Foundation for Basic Research under Nos. 11-519-11-6012, 10-02-90409 and 10-02-00021. Discussions with A. B. Vorontsov, Wei Ku, and Y. K. Bang are appreciated. We would like to thank B. C. Lai, T. N. Dokina, A.A. Virus, K.V. Van and

A. N. Nekrasov for technical support.

**Figure captions**

Fig. 1     $C/T$ vs. $T^2$ with $H$ from 0 to 9 T. The solid line represent the normal state $C_n(T) = \gamma_n T + \beta T^3$ determined from the normal state data of all fields. The inset presents the original $C$ vs. $T$ data taken at $H=0$.

Fig. 2     Superconducting electronic $C_{es}/T$ fit by (a) $d$-wave; (b) extended $s$-wave; (c) Two-gap ($S_\pm$); (d) s+extended $s$-wave models. For the formula of the order parameters in each case, see text. The insets denote the deviations of the fits from the experimental data.

Fig. 3. The comparison of various models focused on the very low temperature regime. The inset shows the entropy conservation required for a second order phase transition and justifies the determination of $C_n(T)$ in Fig. 1. The error bars of data near 0.5 K and 2 K were depicted. The absolute error at low $T$ is about 6% as denoted by the error bar. The relative error can be seen from the slight scattering of the data. (We performed two measurements at each temperature.) The uncertainty from the subtracted phonon term is negligible, since the phonon contribution itself is small at ~0.5 K. On the other hand, The absolute error at $T$~2 K is about 2%, and the uncertainty from the subtracted phonon term contributes about one third of the error

bar.

Fig. 4 (a) The mixed state quasiparticle contribution $\gamma(H)$ for $H//c$ and $H\perp c$ data. The solid lines denote the phenomenological linear fits above $H$=1 T. The dashed line on the top denotes the value of $\gamma_n$. (b) $H_{c2}(T)$ with $H//c$ and $H\perp c$. The inset shows data in the whole range.

| Order parameter | Energy gap(meV) | Weight(%) | $\alpha$ |
|---|---|---|---|
| $d$-wave | $\Delta_0=1.93$ | — | — |
| Extended $s$-wave | $\Delta_e=1.25$ | — | 0.64 |
| two-gap | $\Delta_L=1.55$ | 71 | — |
| | $\Delta_S=0.45$ | 29 | — |
| $s$-wave+ | $\Delta_0=1.33$ | 33 | — |
| Extended $s$-wave | $\Delta_e=1.13$ | 67 | 0.78 |
| $s$-wave+ | $\Delta_0=1.62$ | 63 | $k_r=1.12$ |
| $\Delta_e(\cos k_x+\cos k_y)$ | $\Delta_e=1.49$ | 37 | |

Table I. The parameters derived from the fits in Figs. 2 and 3. For $s$-wave+$\Delta_e(\cos k_x+\cos k_y)$ model, $k_r$ denotes the radius of the electron pockets in the unit of $1/a$ where $a$ is the lattice constant.

| $\gamma_n$ (mJ/mol K$^2$) | $\Theta_D$ (K) | $\delta C/\gamma_n T_c$ | $\lambda$ | $H_{c2,H//c}$ (T) | $H_{c2,H\perp c}$ (T) |
|---|---|---|---|---|---|
| 5.73 | 210 | 1.65 | 1.55 | 13.1 | 27.9 |

Table II. Some fundamental properties of FeSe. $\gamma_n$: normal state electronic coefficient; $\Theta_D$: Debye temperature; $\lambda$: electron-boson coupling constant; $H_{c2,H//c}$: upper critical field for $H//c$; $H_{c2,H\perp c}$: upper critical field for $H\perp c$.

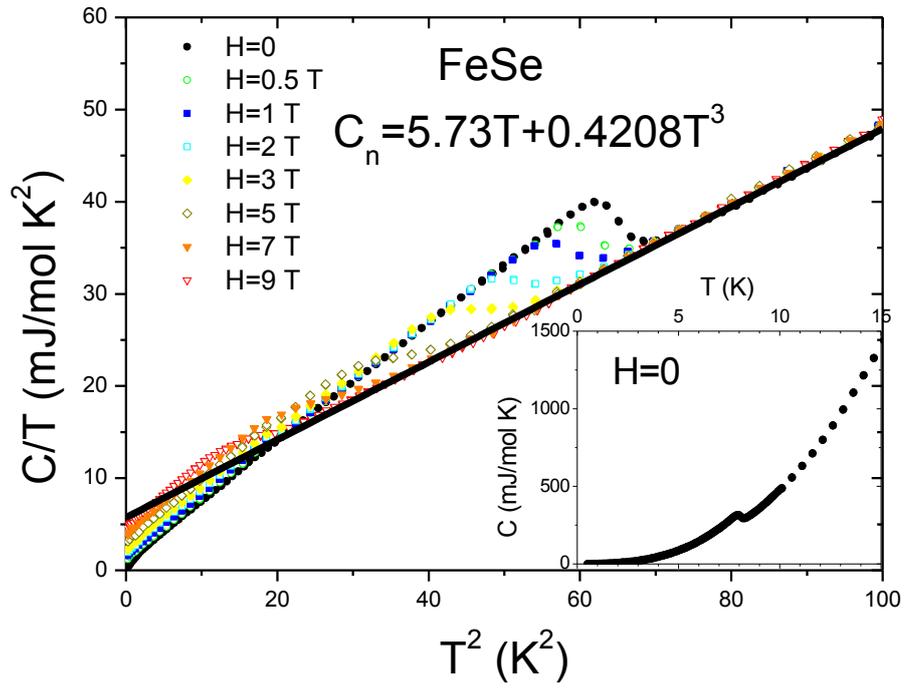

Fig. 1

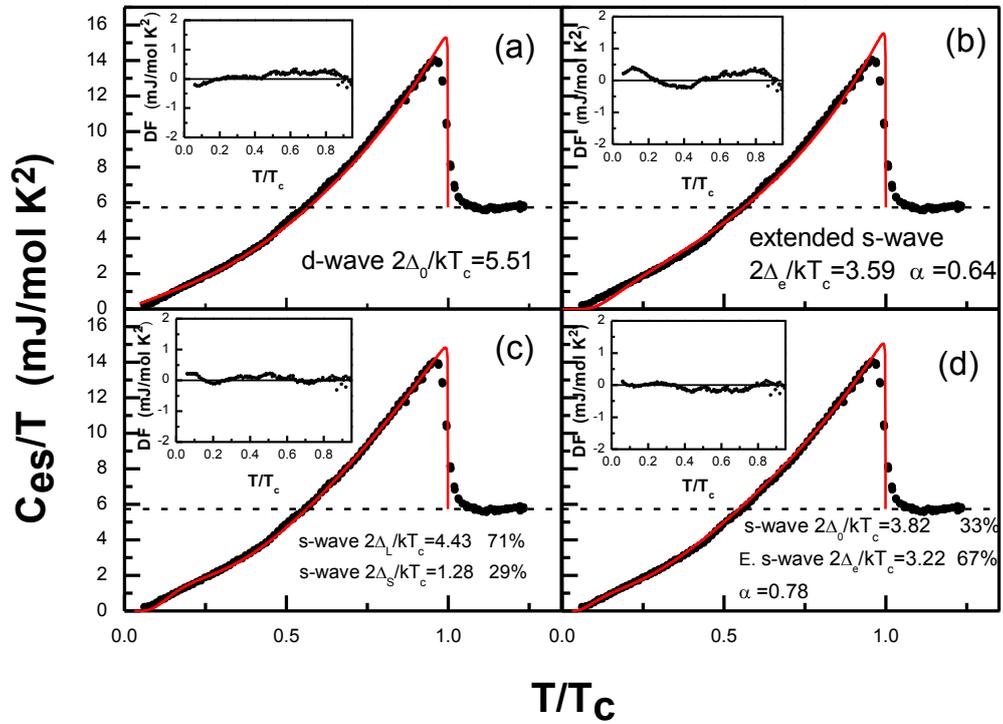

Fig. 2

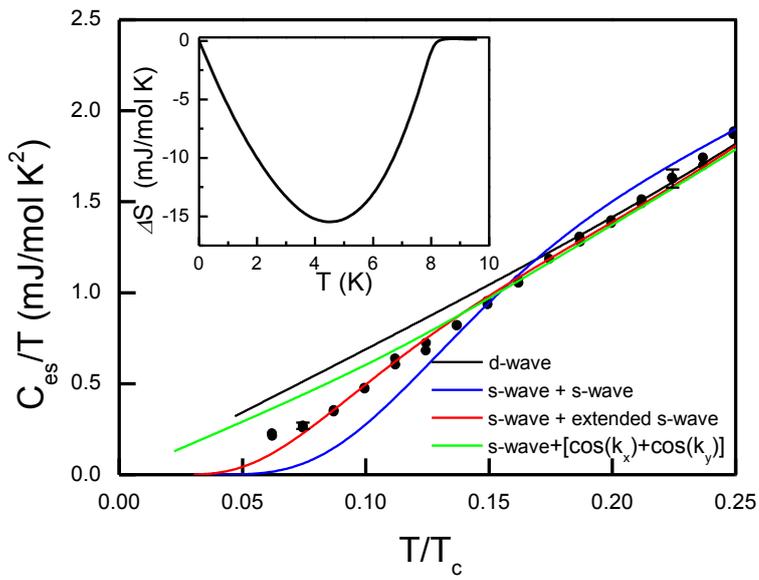

Fig. 3

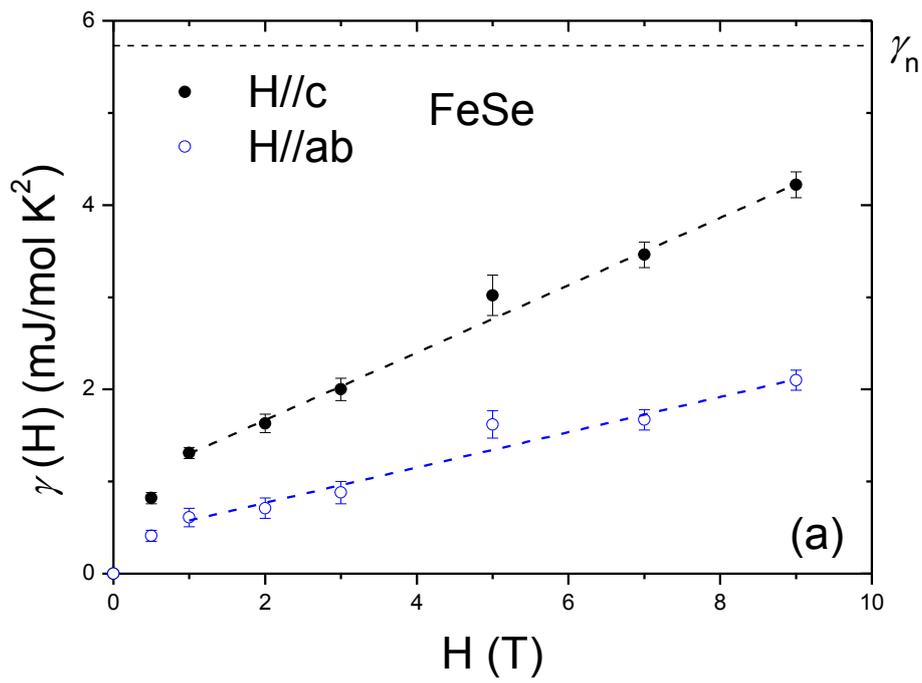

Fig. 4(a)

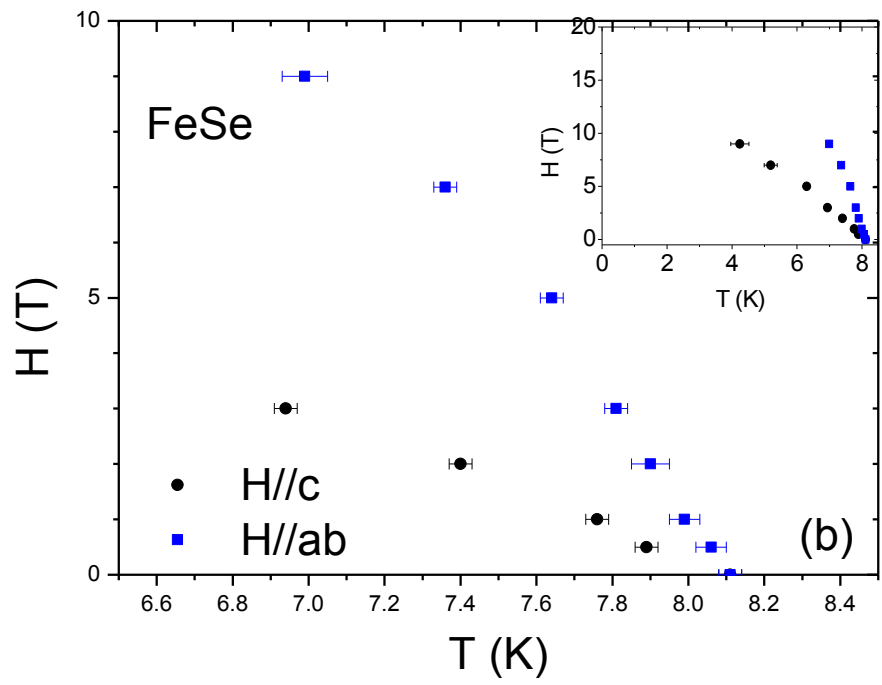

Fig. 4(b)